\documentclass[aps,prb,preprint,superscriptaddress,showpacs,preprintnumbers,amsmath,amssymb]{revtex4-1}
\usepackage{graphicx}

\begin{document}

\preprint{}

\title{"Membrane-outside" as an optomechanical system}

\author{A. K. Tagantsev}
\email{alexander.tagantsev@epfl.ch}
\affiliation{Swiss Federal Institute of Technology (EPFL), School of Engineering, Institute of Materials Science, CH-1015 Lausanne, Switzerland}
\affiliation{Ioffe Phys.-Tech. Institute, 26 Politekhnicheskaya, 194021, St.-Petersburg, Russia}
\author{E. S. Polzik}
\affiliation{Niels Bohr Institute, Quantum Optics Laboratory - QUANTOP, Blegdamsvej 17, DK-2100 Copenhagen, Denmark}
\begin{abstract}
We theoretically study  an optomechanical system, which consists of a two-sided cavity and a mechanical membrane that is placed outside of it.
The membrane is positioned close to one of its mirrors, and the cavity is coupled to the external light field through the other mirror.
Our study is focused on the regime where the dispersive optomechanical coupling in the system vanishes.
Such a regime is found to be possible if the membrane is less reflecting than the adjacent mirror,  yielding a potentially very strong dissipative optomechanical coupling.
Specifically, if the absolute values of amplitude transmission coefficients of the membrane and the mirror, $t$ and $t_m$ respectively, obey the condition $ t_m^2< t\ll t_m\ll 1$,
the \emph{dissipative} coupling constant of the setup exceeds the \emph{dispersive} coupling constant for an optomechanical cavity of the same length.
The dissipative coupling constant and the corresponding optomechanical cooperativity of the proposed system are also compared with those of the Michelson-Sagnac interferometer and the so-called "membrane-at-the-edge" system, which are known for a strong optomechanical dissipative interaction.
It is shown that under the above condition, the system proposed here is advantageous in both aspects.
It also enables an efficient realization of the two-port configuration, which was recently proposed as a promising optomechanical system, providing, among other benefits, a possibility of quantum limited optomechanical measurements in a system, which does not suffer from any optomechanical instability.
 \end{abstract}
\pacs{ 42.50.Lc, 42.50.Wk, 07.10.Cm, 42.50.Ct}

\date{\today}

\maketitle

\newpage

\section{Introduction}

Cavity quantum optomechanics is a promising branch of quantum optics.
It allows for exploration of fundamental issues of quantum mechanics and paves a way for numerous applications, e.g. in high-precision metrology and gravitational-wave defection \cite{Aspelmeyer2014}.
Mainly, the cavity optomechanics profits from the so-called \emph{dispersive coupling}, which originates from the dependence of the cavity resonance frequency on the position of a mechanical oscillator.
However, about a decade ago,  Elste et al~\cite{Elste2009} pointed out that the dispersive coupling does not provide the complete description of the optomechanical interaction.
To fill the gap, those authors have introduced the so-called \emph{dissipative coupling}, which originates from the dependence of the cavity decay rate on the position of the mechanical oscillator.
Since then such a coupling has been attracting an appreciable attention of theorists~\cite{huang2017,Weiss2013,Weiss2013a,Kilda2016,
vyatchanin2016,nazmiev2019,Vostrosablin2014,Tarabrin2013,Xuereb2011,Tagantsev2018,tagantsev2019,mehmood2019,Khalili2016,huang2018,huang2019gen,mehmood2018,dumont2019,tagantsev2020a,tagantsev2020b}
and experimentalists~\cite{Li2009,Sawadsky2015,tsvirkun2015,Wu2014,meyer2016,zhang_2014}.
The dissipative coupling can do virtually all the jobs of the dispersive coupling, such as optomechanical cooling, optical squeezing, and mechanical sensing while the physical conditions and mechanisms encountered in it are rather different.
Among theoretical predictions analyzed for dissipative-coupling-assisted systems are the possibility of simultaneous squeezing and sideband cooling~\cite{Kilda2016}, a stable optical-spring effect, which is not-feed-back-assisted~\cite{nazmiev2019}, a virtually perfect squeezing of the optical noise in a system exhibiting no optomechanical instability\cite{tagantsev2019}, and not-feed-back-assisted cooling of a mechanical oscillator under the resonance excitation~\cite{Tarabrin2013}, the latter also demonstrated experimentally~\cite{Sawadsky2015}.

The experimental implementations of the dissipative coupling are lagging significantly behind the theory.
To a great extend this is due to the fact that the dissipative coupling is typically weak compared to the dispersive coupling.
Hence even under specific tuning conditions when the dispersive coupling vanishes,  it is typically difficult to make the dissipative coupling efficient.
To date, the  Michelson-Sagnac interferometer (MSI) has been theoretically identified~\cite{Xuereb2011} and experimentally addressed~\cite{Sawadsky2015} as a system, which can be tuned to be dominated by an "anomalously strong dissipative coupling".
Recently, the so-called "membrane-at-the-edge" system \cite{dumont2019} (MATE), consisting of a one-sided cavity with a membrane placed inside the main resonator close to the input mirror, has been proposed as a candidate for an enhanced dissipative coupling.
Another system dealing with an anomalously strong dissipative coupling is the popular "membrane-in-the-middle" cavity ~\cite{jayich2008,miao2009,yanay2016,Thompson2008,wilson2009,Purdy2013,mason2019,kampel2017,higginbotham2018} once it is driven close to the point of the spontaneous symmetry breaking~\cite{tagantsev2020a}.
Here even divergence of dissipative coupling constants of individuals modes has been predicted.
However, this system is characterized by tight doublets of modes with opposite signs of the dissipative coupling constants, leading to cancelation of such divergencies and an optomechanical performance very different from that of dissipative coupling of a single mode.

\begin{figure}
\includegraphics [width=0.95\columnwidth] {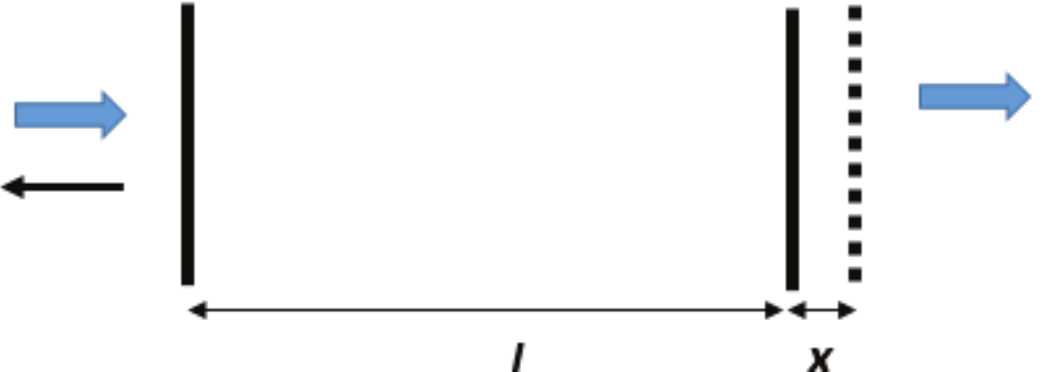}
\caption{Membrane-outside system:  a two-sided cavity with a membrane set behind the second ("non-feeding") mirror. Dashed line - membrane, thin lines -semitransparent mirrors, thick arrows - pumping light, thin arrow - detected light.}
\label{FIG1}
\end{figure}

In the present paper, we treat theoretically an optomechanical system which consists of a two-sided cavity and a membrane that is placed \textit{outside} of it, close to one of its mirrors. The cavity is coupled to the external laser light through the other mirror, and the light leaving the cavity through that mirror is detected (Fig.\ref{FIG1}).
We show that, for a properly positioned membrane which is less reflecting than the adjacent cavity mirror, the dissipative coupling in the system vanishes while the dissipative couling becomes anomalously strong.
We have identified the interval of the membrane transparency providing the superior dissipative optomechanical coupling in this system which we refer to as a "membrane-outside" (MOS).

In addition to an example of an optomechanical device fully controlled by a strong dissipative coupling, MOS enables an efficient realization of the two-port configuration, which was recently proposed\cite{tagantsev2019} as a promising optomechanical system.

We present a theoretical analysis of MOS (Sec.\ref{analysis}), paying special attention to the two-port configuration (Sect.\ref{Two}).
A detailed comparison with MSI (Sec.\ref{MSIcomp}) and MATE (Sec.\ref{MATE}) in terms of the dissipative coupling constant and optomechanical cooperativity in the regimes dominated by the dissipative coupling is provided.
\section{Theoretical analysis of MOS}
\label{analysis}
Primarily, we are interested in finding settings of MOS, under which it does not exhibit the dispersive optomechanical coupling, and in evaluating the strength of the dissipative coupling under these settings.
A simple way to do this is to use the "effective mirror" approach (see, e.g.\cite{Xuereb2011,Sawadsky2015}) following which the tandem  mirror/membrane (Fig.\ref{FIG1}) will be treated as a synthetic mirror.
For a fixed cavity length $l$ any  variation of the mechanical variable, which is the distance $x$ between the membrane and the mirror,  will not affect the cavity  optical length while the cavity decay rate and resonance frequency  will be fully conditioned by the $x$-dependence of the power transmission coefficient of the synthetic mirror and that of the phase of its reflection coefficient, respectively.

We introduce the scattering matrices for the mirror
\begin{equation}
\label{mirror}
 \left(
  \begin{array}{cc}
 it &  -r \\
-r  & it  \\
  \end{array}
\right)\qquad
\end{equation}
and for the membrane
\begin{equation}
\label{membrane}
 \left(
  \begin{array}{cc}
 t_me^{i\varphi_t} &  r_me^{i\varphi_r} \\
r_m e^{i\varphi_r} & t_me^{i\varphi_t}  \\
  \end{array}
\right),
\end{equation}
where $t$ and $r$ are the absolute values of the amplitude transmission and reflection coefficients, respectively, which obeys the following relations
\begin{equation}
\label{parameters}
 t^2+r^2 =1,  \qquad t_m^2+r_m^2 =1, \qquad \mathrm{and}\qquad  e^{2i(\varphi_r-\varphi_t)}=-1.
\end{equation}

Straightforward calculations (see Appendix \ref{ScMa}) yield
\begin{equation}
\label{T}
T = \frac{t^2t_m^2}{1+r^2r_m^2+ 2rr_m \cos\psi},
\end{equation}
for the power transmission coefficient of the synthetic mirror and
\begin{equation}
\label{tanmu}
\tan\mu =r_mt^2\frac{\sin\psi}{r_m(1+r^2)\cos\psi +r(1+r_m^2)}
\end{equation}
for the phase of its the reflection coefficient $\mu$.
Here
\begin{equation}
\label{psi}
\psi=2kx+\varphi_r,
\end{equation}
where $k$ is the light wave vector.

The membrane position where the dispersive coupling vanishes is given by the condition
\begin{equation}
\label{Cond0}
\frac{d\tan \mu}{d\psi}=0.
\end{equation}
Such a derivative reads
\begin{equation}
\label{mux}
\frac{d\tan \mu}{d\psi} =r_mt^2\frac{r_m(1+r^2)+r(1+r_m^2)\cos\psi}{[r_m(1+r^2)\cos\psi +r(1+r_m^2)]^2}.
\end{equation}

Thus, as follows from Eqs.(\ref{Cond0}) and (\ref{mux}), the positions of the membrane where the dispersive coupling vanishes should satisfy the following condition
\begin{equation}
\label{van}
\cos\psi = - \frac{r_m(1+r^2)}{r(1+r_m^2)}.
\end{equation}
This condition can be met if
\begin{equation}
\label{cond}
r_m<r,
\end{equation}
i.e. the membrane should be less reflective than the adjacent mirror.
Thus, under such a condition, at certain values of $\psi$, which is controlled by the position of membrane $x$,  the system will be purely governed by the dissipative coupling, the situation we are looking for.

Let us  check if the positions given by Eq.(\ref{van}) are of practical interest for  implementation in optomechanics.
For this purpose, we will find the range of parameters of  the synthetic mirror where (\ref{van}) is compatible with the basic requirement

\begin{equation}
\label{cond1}
T\ll 1.
\end{equation}
(if any)  and evaluate the dissipative coupling  in this compatible regime.

Inserting (\ref{van}) into (\ref{T}), one finds
\begin{equation}
\label{tau3}
T =t^2 \frac{1+r_m^2}{1-r^2r_m^2}.
\end{equation}
One readily checks that (\ref{cond}) ensures $T\leq 1$ while (\ref{cond1}) and (\ref{van}) are compatible at
\begin{equation}
\label{cond2}
t\ll t_m.
\end{equation}

Now the strength of the dissipative coupling at the point where the dispersive coupling vanishes can be evaluated.
First, if we neglect the energy stored in the synthetic mirror, which is a good approximation for $x\ll l$ (see Appendix \ref{CoupCon}), the decay rate associated with the synthetic mirror  can be written as follows
\begin{equation}
\label{gamma2}
\gamma =\frac{cT}{2l}
\end{equation}
where $c$ is the speed  of light.
Thus, in view of (\ref{psi}), we find
\begin{equation}
\label{der0}
 \frac{d\gamma}{dx}  =\frac{ck}{l} \frac{dT}{d\psi}.
\end{equation}
Next, the following relation
\begin{equation}
\label{T/x}
\frac{dT}{d\psi} = \frac{2rr_m t^2t_m^2\sin\psi}{[1+r^2r_m^2+ 2rr_m \cos\psi]^2}
\end{equation}
and condition (\ref{van}) yield
\begin{equation}
\label{dissConst}
\left|\frac{d\gamma}{dx}\right|= \frac{ck}{l}\frac{t^2}{t_m^2} \frac{2 r_m (1+r_m^2)}{1- r_m^2 r^2}\sqrt{\frac{r^2-r_m^2}{1- r_m^2 r^2}}
\approx \frac{ck}{l}\frac{t^2}{t_m^4}2 r_m (1+r_m^2),
\end{equation}
where condition (\ref{cond2}) was taken into account. This equation predicts a potentially very strong dissipative coupling for $t_m^2<t$.


We conclude that, for the amplitude transparency of the membrane satisfying the condition
\begin{equation}
\label{condTOT}
t_m^2< t\ll t_m\ll 1
\end{equation}
and with the suitably adjusted distance between the membrane and the mirror, MOS can be dominated by the \textit{dissipative} coupling, which is stronger than the \textit{dispersive} coupling $\frac{ck}{l}$ for an optomechanical cavity of the same length.

Equations (\ref{van}), (\ref{condTOT}),  and (\ref{T}) imply that of interest are the positions of the membrane with  $x$ close to
\begin{equation}
\label{x0}
\tilde{x}=\frac{\lambda}{2}\left( \frac{1}{2}+N+\frac{\varphi_r}{2\pi} \right ),\qquad \lambda=2\pi/k,
\end{equation}
where $\cos\psi =-1$ and the transparency of the synthetic mirror is maximal.
We introduce a small parameter
\begin{equation}
\label{Fi}
\Phi = k(x -\tilde{x})\ll 1.
\end{equation}
Keeping the lowest order terms in $\Phi$, one readily finds a set of expressions that describe the system:
\begin{equation}
\label{T1}
T = t^2 \frac{\Phi_0}{\Phi^2+\Phi_0^2},
\end{equation}
\begin{equation}
\label{T/x1}
\frac{dT}{d\Phi} =-2 t^2 \frac{\Phi_0\Phi}{(\Phi^2+\Phi_0^2)^2},
\end{equation}
and
\begin{equation}
\label{mux1}
\frac{d\mu}{d\Phi} = \frac{t^2}{2} \frac{\Phi^2-\Phi_0^2}{(\Phi^2+\Phi_0^2)^2},
\end{equation}
where
\begin{equation}
\label{Fi0}
\Phi_0 = \frac{t_m^2}{4}.
\end{equation}

The above relations enable us to write simple explicit expressions for the cavity decay rate associated with the synthetic mirror
\begin{equation}
\label{gammaEXP}
\gamma= \gamma_0\frac{1}{1+\Phi^2/\Phi_0^2},\qquad \gamma_0=\frac{2c}{l}\frac{t^2}{t_m^2}
\end{equation}
as well as for the optomechanical coupling constants, which we define as follows
\begin{equation}
\label{constants}
g_{\omega0}= -\frac{d\omega_c}{dx}\qquad \mathrm{and} \qquad g_{\gamma0}= -0.5\frac{d\gamma}{dx},
\end{equation}
where $\omega_c$ is a resonance frequency of the system.
Thus, keeping in mind that we are typically interested in $k$ that is very close to $\omega_c/c$, we can write
\begin{equation}
\label{constantsOM}
g_{\omega0}= g_{00} \frac{1-\Phi^2/\Phi_0^2}{(1+\Phi^2/\Phi_0^2)^2}, \qquad g_{\gamma0}=g_{00}\frac{2\Phi/\Phi_0}{(1+\Phi^2/\Phi_0^2)^2},
\qquad g_{00}= \frac{4\omega_c}{l}\frac{t^2}{t_m^4}.
\end{equation}
Here,  when calculating $g_{\omega0}$, we use an approximate relation
\begin{equation}
\label{gomu}
\frac{d\omega_c}{d x}=-\frac{c}{2l}\frac{d\mu}{d x}
\end{equation}
written neglecting the frequency dependence of $\mu$, which is a good approximation for $x\ll l$ (see Appendix \ref{CoupCon}).

\begin{figure}
\includegraphics [width=0.9\columnwidth,clip=true, trim=0mm 0mm 0mm 0mm] {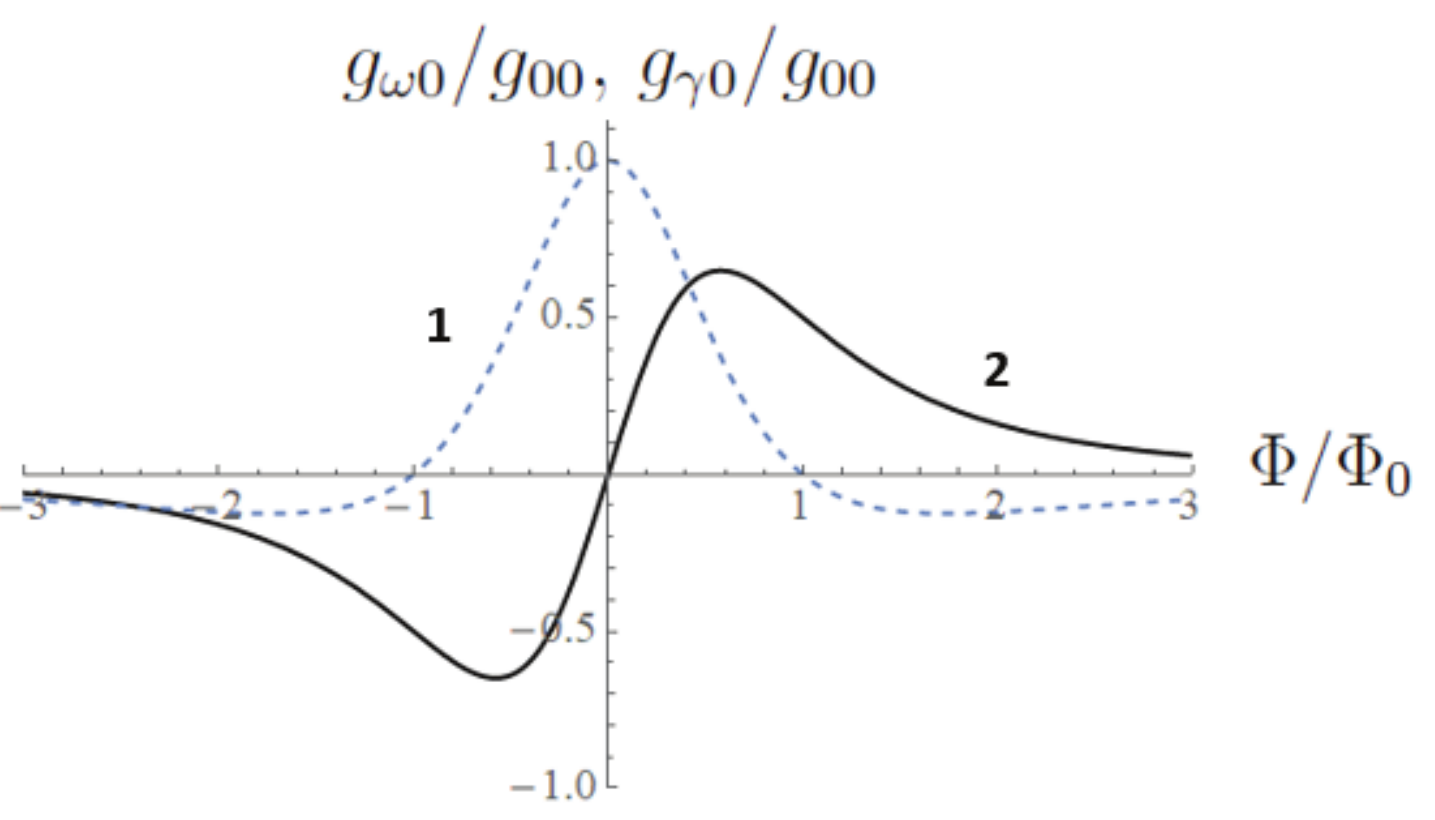}
\caption{Normalized dispersive (1) and dissipative (2) optomechanical constant of MOS,  which are  plotted as functions of $\Phi/\Phi_0 = 4(x-\tilde{x})k/t_m^2$, where $x$ is the  membrane position.
$g_{00}= \frac{4ck}{l}\frac{t^2}{t_m^4}$,
$k$ is the wave vector of the light wave, and $t_m^2$ and $t^2$ are the power transmission coefficients of the membrane and mirror, respectively. $\tilde{x}$ is given by Eq.(\ref{x0}).
\label{FIG2}}
\end{figure}

\begin{figure}
\includegraphics [width=0.9\columnwidth,clip=true, trim=0mm 0mm 0mm 0mm] {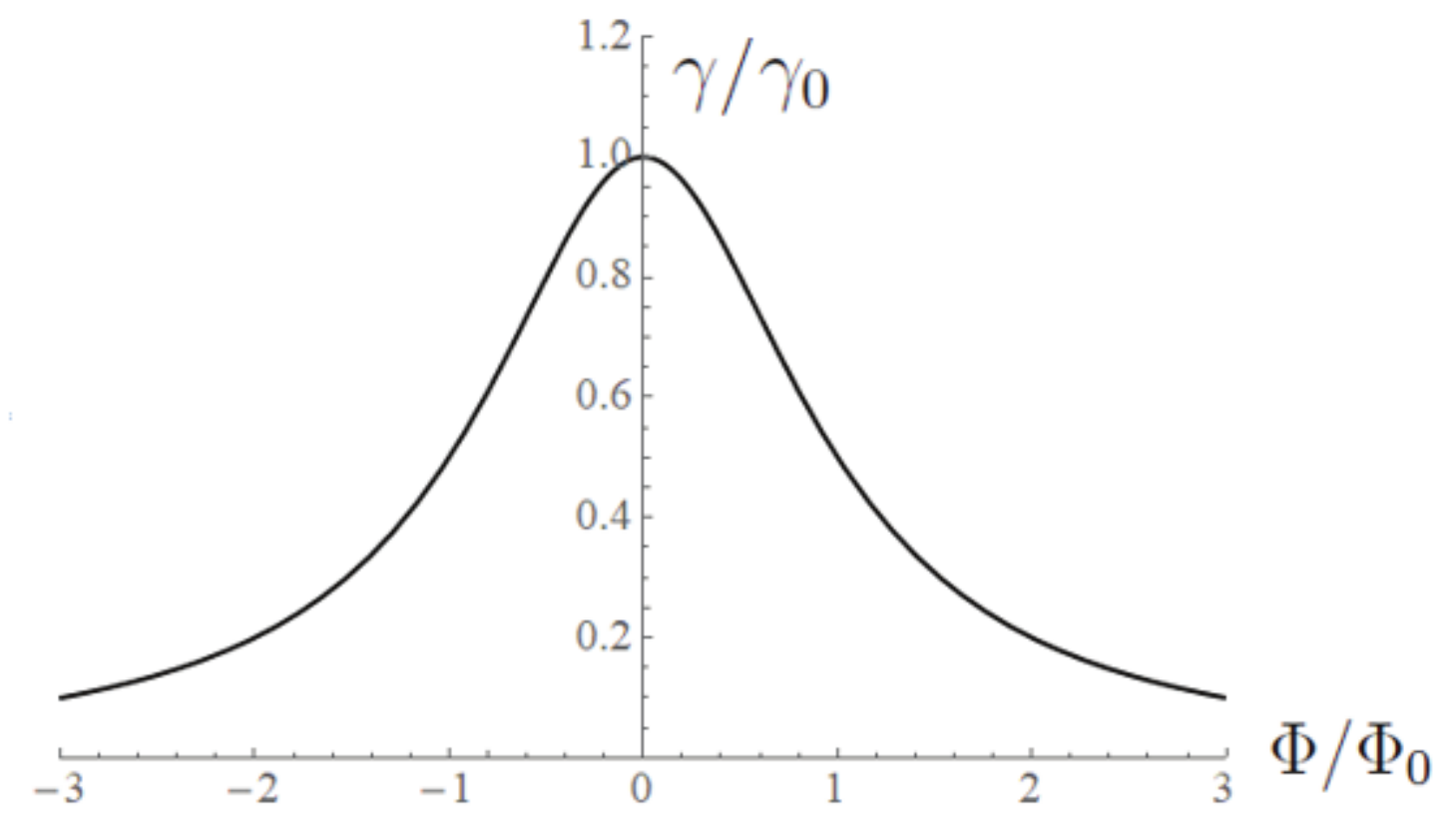}
\caption{Normalized decay rate associated with the synthetic mirror, which is  plotted as a function of $\Phi/\Phi_0 = 4(x-\tilde{x})k/t_m^2$, where $x$ is the  membrane position and $k$ is the wave vector of the light wave.
$\gamma_{0}= \frac{2c}{l}\frac{t^2}{t_m^2}$ and $t_m^2$ and $t^2$ are the power transmission coefficients of the membrane and mirror, respectively. $\tilde{x}$ is given by Eq.(\ref{x0}).
\label{FIG3}}
\end{figure}

The optomechanical constants of the system and the decay rate associated with the synthetic mirror, which are plotted as functions of the mirror position, are shown in Fig.~\ref{FIG2} and Fig.~\ref{FIG3}, respectively.
As illustrated in Fig.~\ref{FIG2}, the position of the reflecting membrane with respect to the adjacent mirrow at which the dissipative coupling is much larger than the dispersive coupling is defined by $\Phi/\Phi_0\approx 1$. This condition imposes the requirement on the membrane position $x -\tilde{x}\approx \lambda t_m^2 /8\pi\approx 0.3$nm which should be maintained with about $20\%$ accuracy. Here we assumed $t_m^2=10^{-2},\lambda=0.85\mu$.

It is also worth elucidating the origin of the enhancement of $\left|\frac{d\gamma}{dx}\right|$ at decreasing $t_m$.
As it is clear from Eqs.(\ref{Fi}),(\ref{T1}), and (\ref{Fi0}), for $t_m\ll 1$ and $x$ close to $\tilde{x}$, the transparency of the synthetic mirror exhibits a sharp maximum, cf., Fig.\ref{FIG3}.
Its height scales as $1/t_m^2$ while its width as $t_m^2$, implying the average slope $ \propto 1/t_m^4$.
It is this  $t_m$-dependence that is seen in Eq.(\ref{dissConst}) for $\left|\frac{d\gamma}{dx}\right|$.

To conclude this Section, we would like to note that strictly speaking, the validity of the presented above results  may require a more stringent condition than  $x\ll l$.
Specifically, as shown in Appendix \ref{CoupCon}, the exact condition reads
\begin{equation}
\label{thickness0}
x\ll l\frac{t_m^4}{4t^2}.
\end{equation}
\section{Implication for symmetric two-sided cavity}
\label{Two}
It was recently shown~\cite{tagantsev2019} that a two-port cavity, which is pumped through one of the mirrors while the transparency of the other composite mirror is modulated with the motion of a mechanical oscillator, is an optomechanical device that is promising for quantum state generation and measurement, not suffering from any optomechanical instability.
For example, it  may be used for quantum limited measurements of the oscillator position and/or for a virtually perfect light squeezing.
This can be realized under the following conditions: (1)~Resonance excitation, (2)~Unresolved side-band regime ("bad cavity limit"), (3)~The system is dominated by the dissipative optomechanical coupling associated with the second port, (4)~The average transparency of the second mirror equals to that of the input mirror ("symmetric" cavity), (5)~The output signal is that reflected from the input mirror.
On the other hand, if a symmetric two-sided optomechanical cavity is dominated by the dispersive coupling, the quantum limit cannot be reached such that  only a 3dB squeezing in possible~\cite{Clerk2010}.

MOS readily enables the realisation of such a device.
For this purpose, one fixes $\Phi=\Phi_0$ by setting the membrane at the distance
\begin{equation}
\label{delx}
\delta x = \lambda \frac{t_m^2}{8\pi}
\end{equation}
from the position $\tilde{x}$ given by Eq.(\ref{x0}) where the synthetic mirror is the most transparent.
Under those conditions the system is governed by the dissipative coupling (Fig.\ref{FIG2}) and the decay rate due to the synthetic mirros is equal to $\gamma_0/2$ (Fig.~\ref{FIG3}).
The decay rate of the input mirror is chosen to be $\gamma_0/2$ to match that of the synthetic mirror.
The cavity is exploited in the unresolved side-band regime.

Remarkably, under such settings, MOS enables switching from the purely dispersive to purely dissipative coupling by a very small displacement of the membrane.
Specifically, as is clear from Fig.\ref{FIG2}, at the $x=\tilde{x}$, i.e. $\Phi=0$, the optomechanical coupling is purely dispersive while after a displacement of the membrane by $\delta x$, given by Eq.(\ref{delx}), it becomes  purely dissipative.
A transition between dissipative and dispersive types of coupling is of special interest since,  according to Ref.~\citenum{tagantsev2019}, it corresponds to the transition between the states of the system where quantum limited measurements are possible and impossible, respectively.
For an ideal situation, where the intracavity losses are absent, such a transition is illustrated by Fig. 2b in Ref.~\citenum{tagantsev2019}.
To estimate the effect of the losses we follow Ref.~\citenum{tagantsev2019} where the backaction-imprecision product $S_{xx}^\textrm{imp} S_{FF}$ was calculated as a function of the ratio of the dispersive to the dissipative coupling constant.
Here $S_{xx}^\textrm{imp}$  is the equivalent displacement noise power spectral density
in the detected light (Fig.\ref{FIG1}) and $S_{FF}$ is the spectral density of the quantum backaction force acting on the membrane. The cavity is driven with a strong monochromatic light.
A quantum limited measurement is possible if $S_{xx}^\textrm{imp} S_{FF}=\hbar^2/4$ where $\hbar$ is the Plank constant.
We generalise the calculations of Ref.~\citenum{tagantsev2019} by incorporating an additional noise source characterized with the decay rate $\gamma_3$ (see Appendix \ref{QuLim}) to find
\begin{equation}
\label{SQLZ}
S_{xx}^\textrm{imp} S_{FF} =\frac{\hbar^2 }{4} \frac{A^2+2A\xi^2}{1+\xi^2},\qquad \xi=\frac{ g_{\omega0}}{ g_{\gamma0}},\qquad A=1+\frac{\gamma_3}{2\gamma}
\end{equation}
for the resonance excitation of the symmetric cavity (the decay rate of both the synthetic and input mirror equals $\gamma$).
\begin{figure}
\includegraphics [width=0.9\columnwidth,clip=true, trim=0mm 0mm 0mm 0mm] {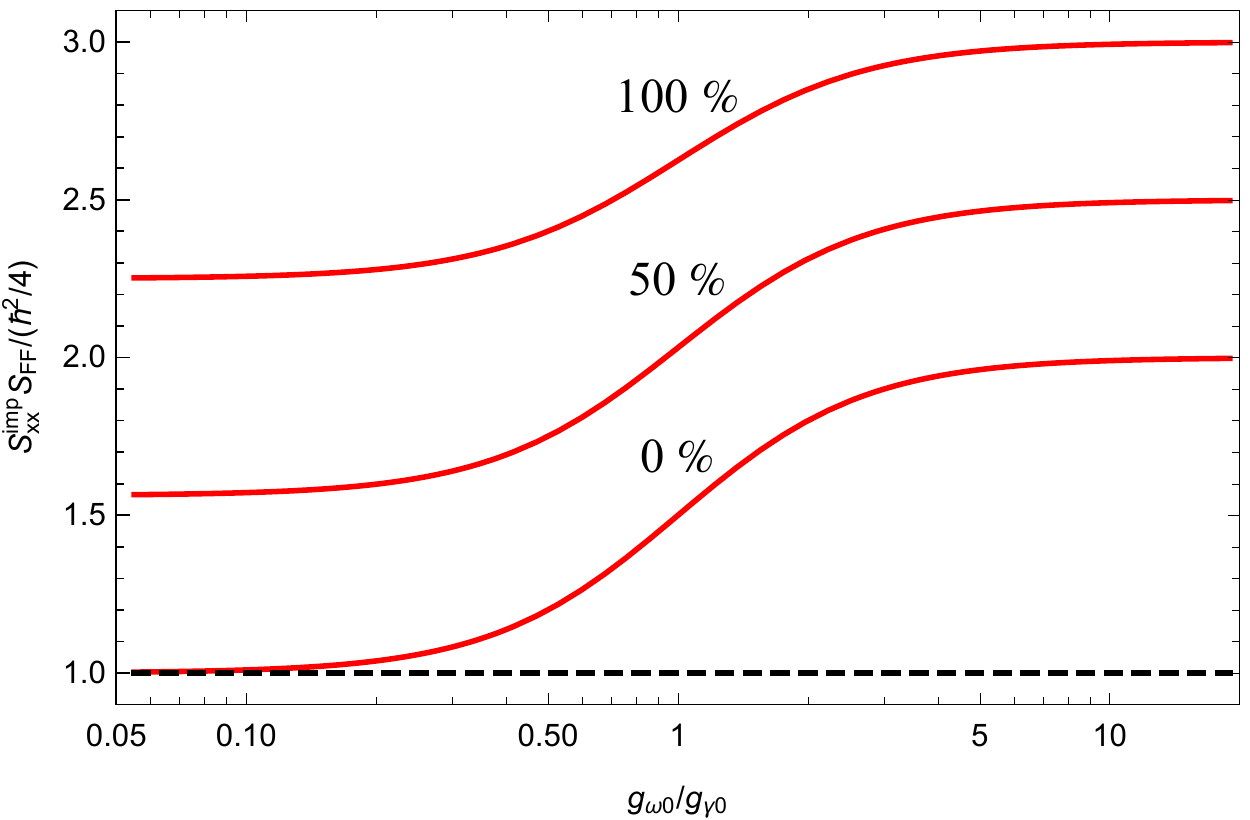}
\caption{The impact of the intracavity losses on backaction-imprecision product for the symmetric membrane-outside system.
 Normalized backaction-imprecision product plotted as a function of the ratio of the optomechanical constants $g_{\omega0}/g_{\gamma0}$ for $\gamma_3=0$ ("0~\%"), $\gamma_3=0.5\gamma$ ("$50\%$"), and $\gamma_3=\gamma$ ("$100\%$") where $\gamma$ is the decay rate through each of the mirrors of the symmetric cavity, and $\gamma_3$ is the decay rate associated with the intracavity losses. We assume the resonance excitation, unresolved sideband regime, and the average transmission of the synthetic mirror being equal to that of the input mirror.
\label{FIG4}}
\end{figure}
Equation (\ref{SQLZ}) is plotted in Fig.\ref{FIG4} for $\gamma_3=0$ ("0~\%"), $\gamma_3=0.5\gamma$ ("$50\%$"), and $\gamma_3=\gamma$ ("$100\%$").
A clear persistence of the kink in this figure suggests that the "switching" effect in question is  rather robust to the presence of the intracavity loss.

The kink shown in Fig.\ref{FIG4} provides a qualitative description of what happens when the membrane is shifted from a position with $\Phi=\Phi_0$ to that with $\Phi=0$.
Quantitatively, the kink is larger because, as follows from Fig.\ref{FIG3}, the shift from $\Phi=\Phi_0$ to $\Phi=0$ also leads to an increase of the decay rate associated with the synthetic mirror, which results in an additional increase of the backaction-imprecision product in the dispersive limit.

Note that the membrane-outside system considered here allows to achieve substantial quantum opto-mechanical cooperativity for the single photon field circulating in the cavity.
Consider the device pumped with a strong monochromatic light ($a_0$ is the number-of-photons-normalized amplitude of the pumping field inside the cavity).
Using the results from  Ref. \citenum{tagantsev2019} for a symmetric two-sided MOS controlled by the dissipative coupling associated with the "non-feeding" mirror, the  cooperativity, via (\ref{gMO}) and (\ref{gammaEXP}), can be expressed as follows
\begin{equation}
\label{CT}
C=
\frac{(g_{\gamma0} x_{\textrm{zpf}}a_0)^2}{\gamma \gamma_m} =M\frac{4t^2}{t_m^6},
\end{equation}
\begin{equation}
\label{M}
M\equiv\frac{c(k a_0 x_{\textrm{zpf}})^2}{l\gamma_m},
\end{equation}
where $x_{\textrm{zpf}}$ is the amplitude of zero-point fluctuations and  $\gamma_m$ is the decay rate of the mechanical oscillator.
For state-of-the-art phononic bandgap membranes \cite{tsaturyan2017} the amplitude of zero-point fluctuations is $x_{\textrm{zpf}}=10^{-15}$m and the mechanical decay rate is $\gamma_m=0.1s^{-1}$. With the cavity length $l=0.1$mm, and the amlplitude transmission coefficients $t_m=0.1 and t=0.014$ for the membrane and the adjacent mirror, respectively, we obtain close to unity cooperativity for a single photon in the cavity ($a_0=1$). The symmetric cavity condition requires that the power transmission coefficient of the input mirror is equal to the effective power transmission coefficient of the synthetic mirror $T=2t^2/t_m^2=0.04$. The corresponding finesse of such symmetric cavity is $F=\pi/T=80$.

\section{Comparison with Michelson-Sagnac interferometer}
\label{MSIcomp}

\begin{figure}
\includegraphics [width=0.7\columnwidth,clip=true, trim=0mm 0mm 0mm 0mm] {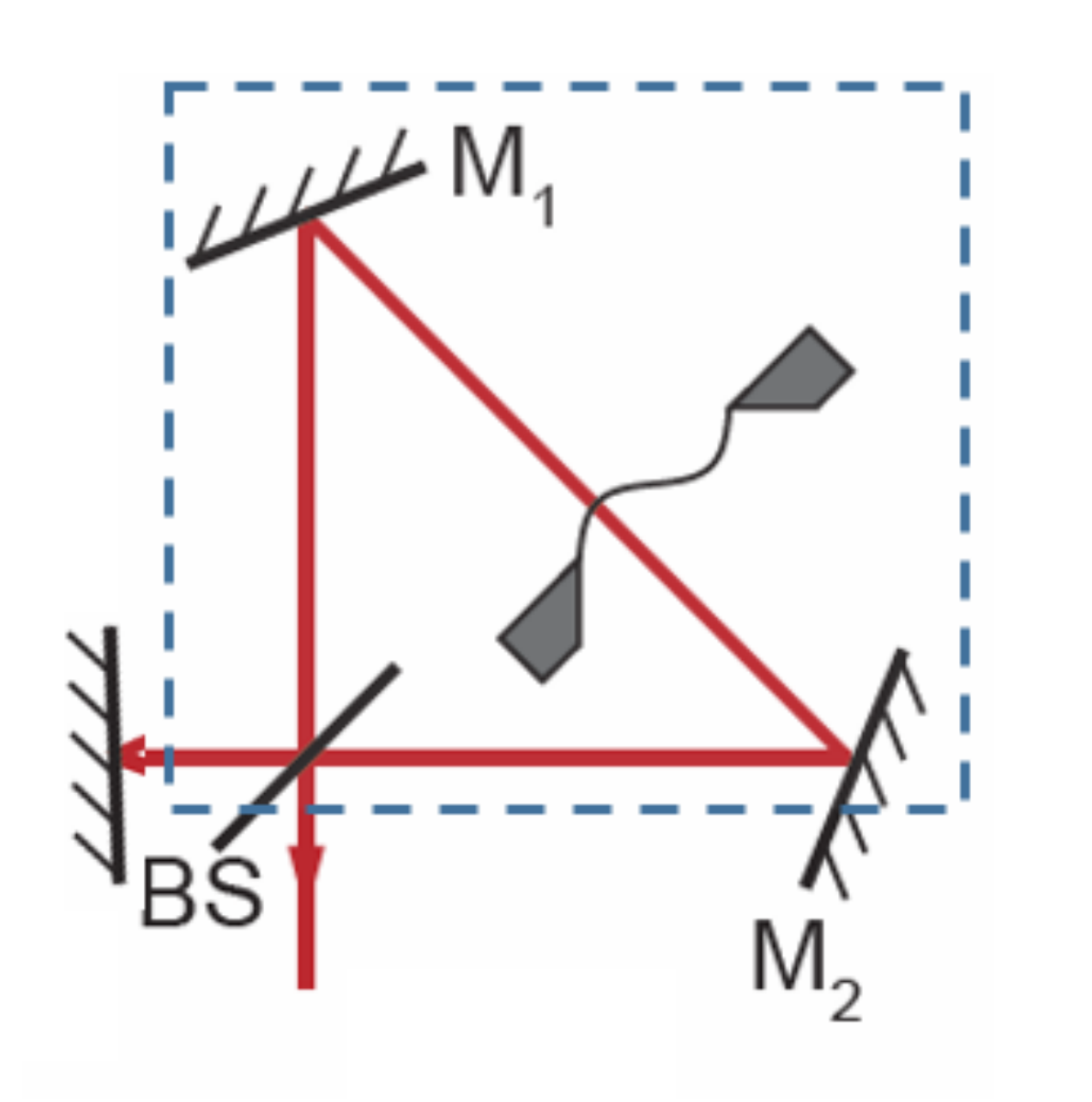}
\caption{Schematic of  Michelson-Sagnac interferometer. The part marked with dashed-line rectangular can be considered as an effective input  mirror with $x$-dependent parameters such that the system can be viewed as a one-sided cavity~\cite{Xuereb2011} .
\label{FIG5}}
\end{figure}

The signal-recycled Michelson-Sagnac interferometer (MSI)~\cite{Xuereb2011,Tarabrin2013} is schematically depicted in Fig.\ref{FIG5}.
It consists of three mirrors, a beam splitter, and a membrane shown with a wiggled line.
This system can be viewed as a one-sided optomechanical cavity with an effective input mirror, the parameters of which are functions of the membrane position~\cite{Xuereb2011}.
For certain  membrane positions the system is controlled exclusively by the dissipative coupling~\cite{Xuereb2011}.
At such positions, in terms of definition (\ref{constants}), the dissipative coupling  constant of MSI can be evaluated as follows (see Appendix~\ref{MSI})
\begin{equation}
\label{gMSI}
|g_{\gamma0}|=r_{\textrm{ms}}\frac{\omega_c}{l}\sqrt{T_{\mathrm{ms}}}
\end{equation}
where $l$ is the effective optical length of the cavity, $r_{\textrm{ms}}$ is the modulus of the amplitude reflection coefficient of the membrane, and $T_{\mathrm{ms}}$ is the power transmission coefficient of the effective mirror.
This result can be compared with the dissipative coupling  constant of MOS at $|\Phi| =\Phi_0$, which, via (\ref{constantsOM}), reads
\begin{equation}
\label{gMO}
|g_{\gamma0}|=2\frac{\omega_c}{l}\frac{t^2}{t_m^4}.
\end{equation}
Clearly for MSI, $|g_{\gamma0}|$ is always appreciably smaller than $\frac{\omega_c}{l}$ while, for MOS, $|g_{\gamma0}|$ can be appreciably larger than $\frac{\omega_c}{l}$.

However, for a balanced comparison, it is reasonable to use the optomechanical cooperativity, which can  serve as a figure of merit for optical squeezing and position measurements.
Consider the device pumped with a strong monochromatic light ($\omega_L$  is  its frequency and $a_0$ is the number-of-photons-normalized amplitude of the pumping field inside the cavity). For MSI as a one-sided cavity in the dissipative coupling regime~\cite{Tagantsev2018}, such a cooperativity, via (\ref{gMSI}), reads
\begin{equation}
\label{CS}
C=
\frac{(g_{\gamma0} x_{\textrm{zpf}}a_0)^2}{\gamma_{\mathrm{ms}}\gamma_m}\left( \frac{2\omega}{\gamma_{\mathrm{ms}}}\right)^2 =2Mr_{\mathrm{ms}}^2\left( \frac{2\omega}{\gamma_{\mathrm{ms}}}\right)^2,
\end{equation}
\begin{equation}
\label{parameters1}
\gamma_{\mathrm{ms}}=\frac{cT_{\mathrm{ms}}}{2l},
\end{equation}
where $M$ comes from (\ref{M}).
Here $\omega = ck- \omega_L$ where $ck$ is the frequency of the detected light.
Typically, $\omega$ is close to the mechanical  resonance frequency $\omega_m$.
Relation (\ref{CS}) is written for the bad cavity regime, i.e. for $\omega_m\ll\gamma_{\mathrm{ms}}$.

Equation (\ref{CS}) is to be compared with the result for MOS given by (\ref{CT}) where, since the dissipative coupling associated with the "non-feeding" mirror,
the small sideband resolve factor $\left( \frac{2\omega}{\gamma_{\mathrm{ms}}}\right)^2$ is absent\cite{tagantsev2019}.
Comparing (\ref{CT}) with (\ref{CS}), a potential advantage of MOS is clearly seen.

\section{Comparison with membrane-at-the-edge system}
\label{MATE}
\begin{figure}
\includegraphics [width=0.7\columnwidth,clip=true, trim=0mm 0mm 0mm 0mm] {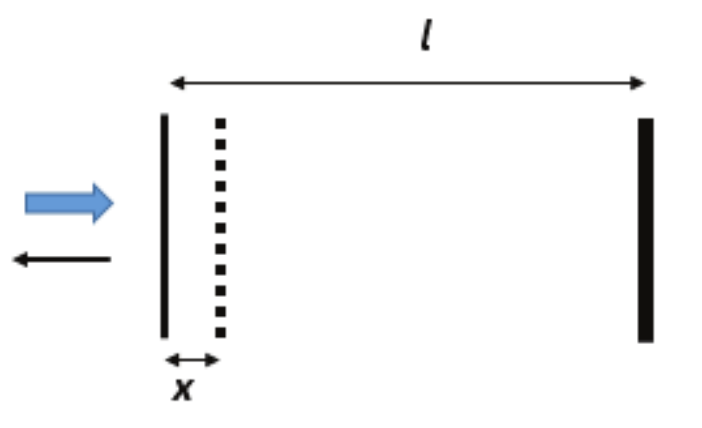}
\caption{Membrane-at-the-edge system: a one-sided cavity with a membrane set inside it, close to the input mirror. Dashed line - membrane, thin solid line - semitransparent input mirror, thick solid line - perfectly reflecting back mirror, thick arrow - pumping light, thin arrow - detected light.
\label{FIG6}}
\end{figure}

The membrane-at-the-edge (MATE) system is a one-sided cavity with a mechanical membrane placed inside it close to the input mirror~\cite{dumont2019} as shown in Fig.\ref{FIG6}.
In Ref.~\citenum{dumont2019}, various optomechanical features of MATE are addressed to demonstrate an advanced optomechanical performance in the case of a highly reflecting membrane.
Specifically, such an advanced performance is identified in a situation where $t_m\ll 1$, $\lambda\ll l$ and
\begin{equation}
\label{x}
x \ll l\frac{t_m^2}{4}
\end{equation}
(see Appendix~\ref{MATEapp}).
Among other features, Ref.~\citenum{dumont2019} covers the dissipative coupling.
The membrane positions where the ratio $|g_{\gamma0}|/\gamma$
is maximal are identified, with $|g_{\gamma0}|$ reaching the value given by Eq.(\ref{gMO}).

This conclusion also readily follows from our results from Sec.~\ref{analysis}.
Indeed, MATE can be viewed as an optomechanical cavity containing the same  synthetic mirror as in MOS, which, however, faces the inner part of the cavity with the opposite side.
The power transmission of such a synthetic mirror is the same in both directions (see Appendix \ref{ScMa}) such that, under condition (\ref{x}),
all results obtained in Sec.~\ref{analysis} for the decay rate and dissipative coupling consonant of MOS hold for MATE system (see Appendix~\ref{MATEapp}).
Next, combining (\ref{constantsOM}) and (\ref{gammaEXP}) we find that $|g_{\gamma0}|/\gamma$  reaches maxima at $\Phi =\Phi_0$, leading to the value of $|g_{\gamma0}|$  given by Eq.~(\ref{gMO}).

At the same time, there is no reason to expect that the dispersive coupling constant of MATE will vanish at $\Phi =\Phi_0$,
since the amplitude reflection coefficients of the synthetic mirror are not the same for the opposite directions (see Appendix \ref{ScMa}) and, in addition, in the case of MATE, the length of the inner part of the cavity is not fixed.
Moreover, as shown in Appendix~\ref{MATEapp}, at $\Phi =\Phi_0$, the system is dominated by the dispersive coupling.

Being interested in the situation where the dispersive coupling is absent, one can show (see Appendix~\ref{MATEapp}) that it occurs at $\Phi^2 =\Phi_0$,
implying, via Eq.~(\ref{constantsOM}),
\begin{equation}
\label{gMATE1}
|g_{\gamma0}| = \frac{\omega_c}{l}\frac {t^2}{t_m},
\end{equation}
for $t_m\ll 1$.
Thus, comparing this result with (\ref{gMO}), one concludes that, in terms of the dissipative coupling constant,  MATE is less advantageous than MOS.

The same conclusion holds in terms of the cooperativity.
Indeed, at $\Phi^2 =\Phi_0$, Eq.~(\ref{gammaEXP}) yields
\begin{equation}
\label{MATEdecay}
\gamma_{\mathrm{\textrm{mate}}} = \frac{ct^2}{2l},
\end{equation}
for the MATE decay rate, implying
\begin{equation}
\label{CMATE}
C=
\frac{(g_{\gamma0} x_{\textrm{zpf}}a_0)^2}{\gamma_{\mathrm{mate}}\gamma_m}\left( \frac{2\omega}{\gamma_{\mathrm{mate}}}\right)^2 =M\frac{t^2}{t_m^2}\left( \frac{2\omega}{\gamma_{\mathrm{mate}}}\right)^2
\end{equation}
for the optomechanical cooperativity in the bad cavity regime.
A comparison of this result with (\ref{CT}) is clearly in favour of MOS.
\section{Conclusions}
\label{Conclusions}
We have theoretically addressed  an optomechanical system which consists of a two-sided cavity and a membrane that is placed outside of it, close to one of its mirrors, while the cavity is fed from the other mirror, and the light leaving it through this mirror being detected.
We term such a setup as membrane-outside system (MOS).
We have shown that, if the membrane is less reflecting than the adjacent mirror and it is positioned very close the point $\tilde{x}$ where the transparency of the mirror/membrane tandem is maximal, the dispersive coupling can be fully suppressed while the dissipative coupling constant can be potentially record high.
Specifically, if
\begin{equation}
\label{condTOT1}
t_m^2 < t\ll t_m\ll 1,
\end{equation}
where $t$ and $t_m$ are the absolute values of amplitude transmission coefficients of the membrane and the mirror, respectively, and the membrane is displaced from $\tilde{x}$ by
\begin{equation}
\label{delx1}
\delta x = \lambda \frac{t_m^2}{8\pi}
\end{equation}
the system is governed by the \emph{dissipative} optomechanical interaction with the coupling constant, which  exceeds the \emph{dispersive} coupling constant for an optomechanical cavity of the same length.

MOS enables an efficient realization of the two-port configuration, which was recently proposed\cite{tagantsev2019} as a promising optomechanical system, allowing among other benefits, e.g., a possibility of quantum limited optomechanical measurements in a system, which does not suffer form any optomechanical instability.
 Such a setup also enables a kind of switching between the regimes where the quantum limited optomechanical measurements are possible and where they are not.
 It is shown that manifestation of that switching is robust to the presence of an appreciable intracavity loss.

 The optomechanical performance of MOS is compared with that of other systems, where the dissipative coupling is viewed as strong: with the Michelson-Sagnac interferometer (MSI)~\cite{Xuereb2011,Sawadsky2015,Tarabrin2013} and with the so-called "membrane-at-the-edge" system (MATE)~\cite{dumont2019}.
 This comparison is performed in terms of the dissipative coupling constant and optomechanical cooperativity for the regime where the dispersive coupling is absent.
 It is found that, for an optimised set of these parameters, the optomechanical performance of MOS is advantageous in both aspects.

 All in all we have identified a system, which, among all the systems dominated by the dissipative optomechanical coupling, exhibits the strongest optomechanical interaction.

 \section{Acknowledgements}
 ESP acknowledges the support of Villum Investigator grant no. 25880, from the Villum Foundation and the ERC Advanced grant QUANTUM-N, project 787520.

\appendix
\section{The scattering matrix of the synthetic mirror}
\label{ScMa}
\begin{figure}
\includegraphics [width=0.7\columnwidth,clip=true, trim=0mm 0mm 0mm 0mm] {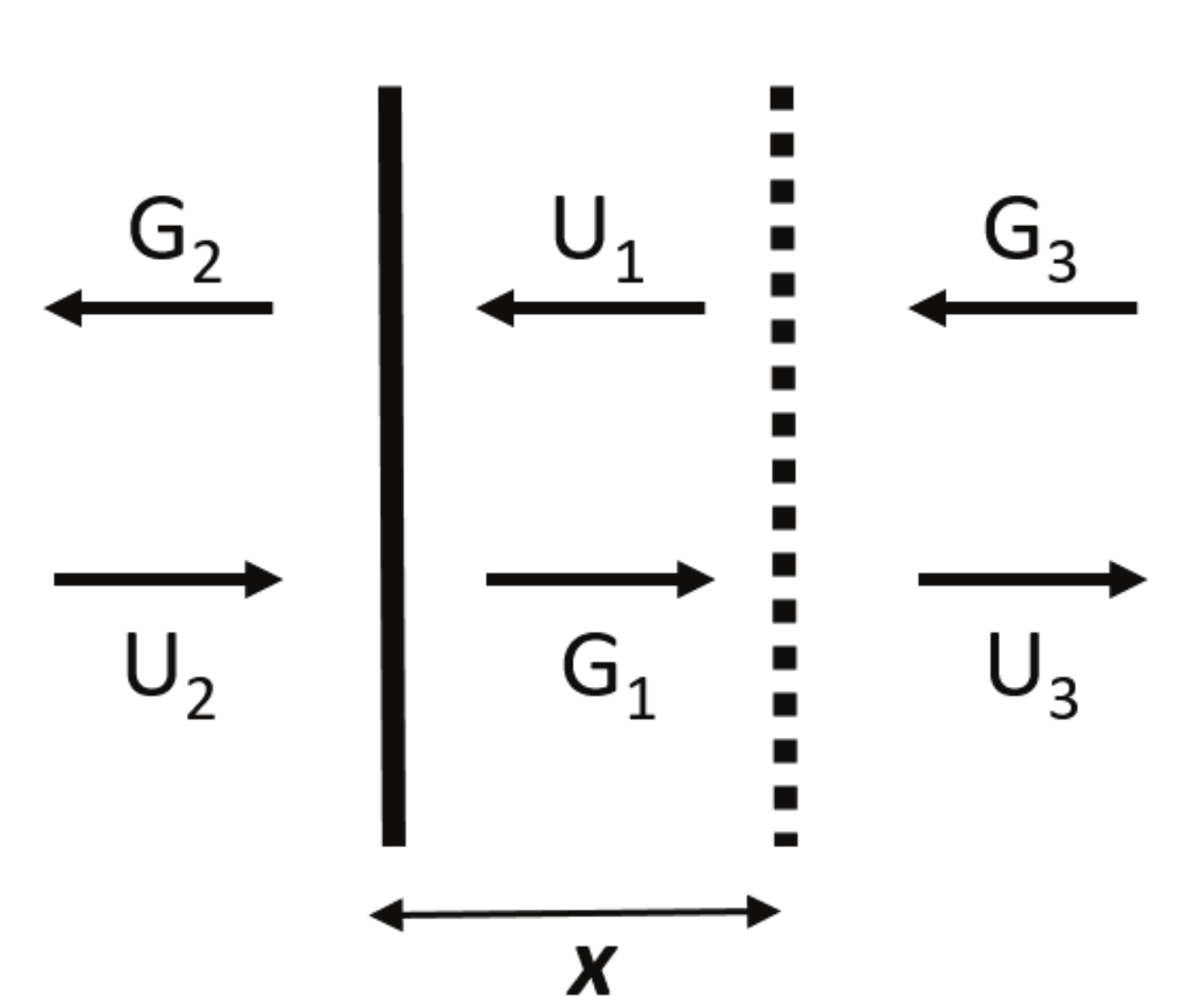}
\caption{A synthetic mirror consisting of a semitransparent mirror (solid line) and  a semitransparent membrane (dashed line).
Running electromagnetic waves are schematically shown with arrows and labeled  with their complex amplitudes.
\label{FIG7}}
\end{figure}

The synthetic mirror in question is schematically depicted in Fig.\ref{FIG7}.
It consists of a semitransparent mirror shown with the solid line and a semitransparent membrane shown with the dashed line.
Their scattering parameters are given by Eqs.~(\ref{mirror}) and (\ref{membrane}).
The complex amplitudes of the wave $G_1$, $G_2$, $G_3$, $U_1$, $U_2$, and $U_3$, which are shown in Fig.\ref{FIG7}, are linked by the following relations
\begin{align}
\begin{aligned}
\\&G_1=itU_{2} - r U_{1},
\\&G_{2}=-r U_{2}+itU_{1},
\\&U_3=t_me^{i\varphi_t}G_{1} + r_me^{i\varphi_r-2ikx} G_{3},
\\&U_{1}e^{-ikx}=r_me^{i\varphi_r+2ikx} G_{1}+t_me^{i\varphi_t}G_{3},
\end{aligned}
\label{set}
\end{align}
where all amplitudes are taken at the  mirror.
The wave vector of the light is denoted as $k$.

We are looking for the scattering matrix $\mathbb{M}$ of the whole system, which is defined as follows
\begin{equation}
\label{definition}
\left(
  \begin{array}{cc}
U_3 \\
G_2  \\
  \end{array}
\right)=\mathbb{M}
 \left(
  \begin{array}{cc}
U_2 \\
G_3  \\
  \end{array}
\right).
\end{equation}
Equations (\ref{set}) readily imply
\begin{equation}
\label{Matrix}
\mathbb{M}= \frac{1}{1+rr_me^{i\psi}}\left(
  \begin{array}{cc}
itt_me^{i\varphi_t} & e^{2i\varphi_r}(r+r_me^{-i\psi}) \\
-r-r_me^{i\psi} &itt_me^{i\varphi_t} \\
  \end{array}
\right),
\end{equation}
where $\psi=2kx+\varphi_r$.
\section{Applicability of the synthetic mirror approach to MOS}
\label{CoupCon}
Confider a two-sided cavity of a fixed length $l$, where  one of the mirrors is reflecting with the $\pi$ phase shift while the other one is synthetic.
We would like to find out how thin the tandem mirror/membrane should be to justify the applicability of the above synthetic mirror approach.
\subsection{Dispersive coupling constant}
One readily checks that the resonance frequencies of the system $\omega_c$ are equal to $ck_c$, where $k_c$ satisfies the following equation
\begin{equation}
\label{equ}
2lk =\pi+2\pi N  - \mu (kx).
\end{equation}
Here $N$ is integer and $\mu(kx)$ is the phase shift at the synthetic mirror.
In view of Eqs.~(\ref{tanmu}) and (\ref{psi}), $\mu$ a function of $kx$.
Equation (\ref{equ}) implies
\begin{equation}
\label{go}
\frac{d\omega_c}{dx} =-\frac{\omega_c\mu'}{2l}\frac{1}{1+\frac{x\mu'}{2l}},
\end{equation}
$$
\mu'=\frac{d\mu}{d(kx)}.
$$
Taking into account that, according to Eq.~(\ref{mux1}) in the range of interest, $|\mu'|$ is about $8t^2/t_m^4$ or smaller, we conclude that, if
\begin{equation}
\label{thickness}
x\ll l\frac{t_m^4}{4t^2},
\end{equation}
the second fraction in (\ref{go}) can be replaced with $1$ to yield
\begin{equation}
\label{thickness1}
\frac{d\omega_c}{d x}=-\frac{c}{2l}\frac{d\mu}{d x}.
\end{equation}
Equations (\ref{thickness}) and ~(\ref{thickness1}) bring us to Eqs.~(\ref{thickness0}) and (\ref{gomu}) from the main text.

\subsection{Decay rate and dispersive coupling constant}
In terms of the complex amplitudes (see Fig.~\ref{FIG7}), the decay rate associated with the synthetic mirror can be written as follows~\cite{dumont2019}
\begin{equation}
\label{gammaNEW}
\gamma =\frac{\dot{W}}{W} = \frac{ct_m^2|G_1|^2}{2(l|U_2|^2+x|G_1|^2)}=\frac{ct_m^2}{2(l|U_2|^2/|G_1|^2+x)},
\end{equation}
where  $W$ is the energy stored in the system and $\dot{W}$ is the dissipated power.
Equations (\ref{set}) and (\ref{T}) imply
\begin{equation}
\label{ratioUG}
\frac{|U_2|^2}{|G_1|^2}= \frac{1+r^2r_m^2+ 2rr_m \cos\psi}{t^2}=\frac{t_m^2}{T}
\end{equation}
such that Eq.~(\ref{gammaNEW}) can be rewritten as follows
\begin{equation}
\label{gammaNEW1}
\gamma = \frac{cT}{2l}\frac{1}{1-\frac{x}{l}\frac{T}{t_m^2}}.
\end{equation}
Taking into account that, in the case of interest, $T$  is about $4t^2/t_m^2$ or smaller(see Eq.~(\ref{T1})), we conclude that, if $x$ is small enough such that inequality (\ref{thickness})  is satisfied, the second fraction in (\ref{gammaNEW1}) can be replaced with $1$ to justify Eq.(\ref{gamma2}) from the main text.

Next, Eq.~(\ref{gammaNEW1}) yields
\begin{equation}
\label{gamma2/x}
\frac{d\gamma}{dx} =ct_m^2\frac{1+\frac{lt_m^2}{T^2}\frac{dT}{dx}}{2(lt_m^2/T+x)^2}.
\end{equation}
Using (\ref{T1}) and (\ref{T/x1}), in the case of interest, $\frac{lt_m^2}{T^2}\frac{dT}{dx}$ can be evaluated as $4kl/T \gg1$ such that $1$ in the numerator in (\ref{gamma2/x}) can be neglected.
As a result, one concludes that, if $x$ is small enough such that inequality (\ref{thickness})  is satisfied, Eq.~(\ref{gamma2/x}) can be rewritten as follows $\frac{d\gamma}{dx} =\frac{c}{2l}\frac{dT}{dx}$, justifying the calculation of the dissipative coupling constant by using the synthetic mirror approach.

\section{The backaction-imperfection product in the presence of intracavity losses}
\label{QuLim}

To evaluate the impact of the intracavity losses on the backaction-imperfection product of a two-port cavity, we model the intracavity losses  as the third port.
The system is pumped with a strong coherent light of frequency $\omega_L$ from the first port, the light backscattered from this port is detected.
We describe the fluctuations in the system with the following equations written for the Fourier transforms of all variables (the argument $\omega$ is dropped) in the reference rotating with frequency $\omega_L$:
\begin{equation}
\label{X}
\left[\frac{\gamma_1+\gamma_2+\gamma_3}{2}-i\omega\right]\textbf{X}+\Delta\textbf{Y}
=\frac{\sqrt{\gamma_1}}{2}\textbf{X}_{\textrm{in1}}+
\frac{\sqrt{\gamma_2}}{2}\textbf{X}_{\textrm{in2}}+\frac{\sqrt{\gamma_3}}{2}\textbf{X}_{\textrm{in3}}
+a_0g_{\gamma0}\textbf{x},
\end{equation}
\begin{equation}
\label{Y}
\left[\frac{\gamma_1+\gamma_2+\gamma_3}{2}-i\omega\right]\textbf{Y}-\Delta\textbf{X}
=\frac{\sqrt{\gamma_1}}{2}\textbf{Y}_{\textrm{in1}}+\frac{\sqrt{\gamma_2}}{2}\textbf{Y}_{\textrm{in2}}+
\frac{\sqrt{\gamma_3}}{2}\textbf{X}_{\textrm{in3}}+a_0g_{\omega0}\textbf{x},
\end{equation}
\begin{equation}
\label{F}
\textbf{F}=-a_0\frac{\hbar g_{\gamma0}}{\sqrt{\gamma}}\textbf{Y}_{\textrm{in2}}+2a_0\hbar g_{\omega0}\textbf{X}.
\end{equation}
where $g_{\gamma0}$ and $g_{\omega0}$ are defined by (\ref{constants}).
Here $\Delta=\omega_L-\omega_c$, where $\omega_c$ is the resonance frequency, $\gamma_{1,2,3}$ are the decay rates of the three ports, $a_0$ is a number-of-photon normalized amplitude of the intracavity pumping field.
The operator of mechanical displacement is denoted as $\mathbf{x}$.
The quadratures of operators of fluctuating parts of the intracavity field, $\mathbf{a}$, and those of the input fields, $\textbf{A}_{\textrm{in1,2,3}}$,  are defined as follows
$$\textbf{X}(\omega)=[\textbf{a}(\omega)+\textbf{a}^\dag(-\omega)]/2,$$ $$\textbf{Y}(\omega)=-i[\textbf{a}(\omega)-\textbf{a}^\dag(-\omega)]/2,$$
$$\textbf{X}_{\textrm{in1,2,3}}(\omega)=\textbf{A}_{\textrm{in1,2,3}}(\omega)+\textbf{A}_{\textrm{in1,2,3}}^\dag(-\omega),$$
$$\textbf{Y}_{\textrm{in1,2,3}}(\omega)=-i[\textbf{A}_{\textrm{in1,2,3}}(\omega)-\textbf{A}_{\textrm{in1,2,3}}^\dag(-\omega)].$$
The correlators of the field quadratures satisfy the following relations
\begin{align}
\begin{aligned}
&\langle\textbf{X}_{\textrm{in1,2,3}}(\omega)\textbf{X}_{\textrm{in1,2,3}}(\omega')\rangle=
\langle\textbf{Y}_{\textrm{in1,2,3}}(\omega)\textbf{Y}_{\textrm{in1,2,3}}(\omega')\rangle
=
\\&i\langle\textbf{Y}_{\textrm{in1,2,3}}(\omega)\textbf{X}_{\textrm{in1,2,3}}(\omega')\rangle=
-i\langle\textbf{X}_{\textrm{in1,2,3}}(\omega)\textbf{Y}_{\textrm{in1,2,3}}(\omega')\rangle
\\&=\delta(\omega+\omega'),
\end{aligned}
\label{XY1}
\end{align}
where $\langle ... \rangle$ stands for the ensemble averaging.

The output field from the first port, which is detected, obeys the following relation
\begin{equation}
\label{IOXO}
\textbf{X}_{\textrm{in}1}+\textbf{X}_{\textrm{out}1}=2\sqrt{\gamma_1}\textbf{X},\qquad \textbf{Y}_{\textrm{in}1}+\textbf{Y}_{\textrm{out}1}=2\sqrt{\gamma_1}\textbf{Y}.
\end{equation}

We are interested in the backaction-imperfection product for the symmetric two-sided cavity ($\gamma_1=\gamma_2=\gamma$),  the resonance excitation ($\Delta=0$), and in  the low frequency limit ($\omega/\gamma\Rightarrow0$).
In such a situation, using the above relations, we find
\begin{equation}
\label{OUT1X}
\textbf{X}_{\textrm{out}1} =\tilde{\textbf{X}}_{\textrm{in}} +
\frac{2 a_0g_{\gamma0}\sqrt{\gamma}}{\gamma+\gamma_3/2}\textbf{x},
\end{equation}
\begin{equation}
\label{OUT1Y}
\textbf{Y}_{\textrm{out}1} =\tilde{\textbf{Y}}_{\textrm{in}} +\frac{2 a_0g_{\omega0}\sqrt{\gamma}}{\gamma+\gamma_3/2}\textbf{x},
\end{equation}
where the input noise operators $\tilde{\textbf{X}}_{\textrm{in}}$ and $\tilde{\textbf{Y}}_{\textrm{in}}$ evidently meet Eqs.~(\ref{XY1}).

The optimal quantum-mechanical measurements must employ the  quadrature
$
\textbf{Z}_{\textrm{out}}=\textbf{X}_{\textrm{out1}}\cos\theta+\textbf{Y}_{\textrm{out1}}\sin\theta
$
such that the orthogonal quadrature carries no information about $\textbf{x}$.
This condition is met at $\theta =\tan^{-1}( g_{\omega0}/ g_{\gamma0})$.
For the optimal quadrature, we find
\begin{equation}
\label{OUT1Z}
\textbf{Z}_{\textrm{out}} =\textbf{Z}_{\textrm{in}} +a_0
\frac{2\sqrt{\gamma}\sqrt{ g_{\gamma0}^2+ g_{\omega0}^2}}{\gamma+\gamma_3/2}\textbf{x},
\end{equation}
where the input noise operator $\textbf{Z}_{\textrm{in}}$ obeys  relations (\ref{XY1}),
implying the following spectral power density for the imprecision of position measurements
\begin{equation}
\label{SxxZ}
S_{xx}^\textrm{imp} =\frac{(\gamma+\gamma_3/2)^2}{4a_0^2\gamma}\frac{1}{ g_{\gamma0}^2+ g_{\omega0}^2}.
\end{equation}

In the situation considered, for the stochastic backaction force,  Eqs.~(\ref{X}) and (\ref{F}) yield
\begin{equation}
\label{F1AD}
\textbf{F}=
-\frac{\hbar a_0}{\sqrt{\gamma}} g_{\gamma0}\textbf{Y}_{\textrm{in2}}+ \hbar a_0g_{\omega0}\frac{\sqrt{\gamma}}{\gamma+\gamma_3/2}(\textbf{X}_{\textrm{in1}}
+\textbf{X}_{\textrm{in2}}+\textbf{X}_{\textrm{in3}}\sqrt{\gamma_3/\gamma}),
\end{equation}
which, via (\ref{XY1}), leads to the following expression for the  spectral density of this force
\begin{equation}
\label{SFFZ}
S_{FF} =\frac{\hbar^2 a_0^2\gamma}{(\gamma+\gamma_3/2)^2}\left[\left(1+\frac{\gamma_3}{2\gamma} \right)^2 g_{\gamma0}^2+2\left(1+\frac{\gamma_3}{2\gamma} \right) g_{\omega0}^2 \right].
\end{equation}
Combining (\ref{SxxZ}) and (\ref{SFFZ}), we arrive at the following  backaction-imperfection product
\begin{equation}
\label{SQLZ1}
S_{xx}^\textrm{imp} S_{FF} =\frac{\hbar^2 }{4} \frac{A^2+2A\xi^2}{1+\xi^2},\qquad \xi=\frac{ g_{\omega0}}{ g_{\gamma0}},\qquad A=1+\frac{\gamma_3}{2\gamma},
\end{equation}
which is given by Eq.(\ref{SQLZ}) of the main text.
\section{Michelson-Sagnac interferometer}
\label{MSI}
The Michelson-Sagnac interferometer (MSI) is schematically depicted in Fig.~\ref{FIG5}.
It consists of a beam splitter, a membrane and three perfectly reflecting mirrors.
The beam splitter and the membrane are characterized by following scatting matrices
\begin{equation}
\label{BS}
 \left(
  \begin{array}{cc}
T_b &  -R_b \\
R_b  &  T_b  \\
  \end{array}
\right)
\qquad \textrm{and}
\qquad
 \left(
  \begin{array}{cc}
-r_{\mathrm{ms}} &  t_{\mathrm{ms}} \\
t_{\mathrm{ms}}  &  r_{\mathrm{ms}} \\
  \end{array}
\right),
\end{equation}
respectively, where all coefficients of the matrices are real and positive; $t_{\mathrm{ms}}$ and $T_b$ stand for the amplitude transmission coefficients.
The membrane is displaced to the left from its symmetric position  by the distance $x$.
According to Ref.\citenum{Tarabrin2013}, MSI  can be treated as  an optomechanical cavity of a fixed length $l$  with the input mirror, the scattering matrix of which reads~\cite{Tarabrin2013}
\begin{equation}
\label{Matrix1}
\mathbb{M}= \left(
  \begin{array}{cc}
\rho & \tau \\
\tau &-\rho^* \\
  \end{array}
\right), \qquad \rho=|\rho| e^{i\mu},
\end{equation}
\begin{equation}
\label{MSrho}
\rho=  -2R_bT_bt_{\mathrm{ms}} -(R_b^2-T_b^2) r_{\mathrm{ms}} \cos2kx + ir_{\mathrm{ms}} \sin2kx,
\end{equation}
\begin{equation}
\label{MStau}
\tau= t_{\mathrm{ms}}(T_b^2-R_b^2)+2R_bT_br_{\mathrm{ms}} \cos2kx,
\end{equation}
where $\tau$ stands for the amplitude transmission coefficient, while, for the decay rate and the optomechanical coupling constants, the following relations can be used:
\begin{equation}
\label{MSgamma1}
\gamma_{\mathrm{ms}}=\frac{cT_{\mathrm{ms}}}{2l},\qquad T_{\mathrm{ms}}=\tau^2
\end{equation}
for the decay rate and
\begin{equation}
\label{go1}
g_{\omega0}=-\frac{d\omega_c}{dx} =\frac{d\mu}{dx}
\frac{ c}{2l},
\end{equation}
\begin{equation}
\label{gg1}
g_{\gamma0}=-\frac{1}{2}\frac{d\gamma_{\mathrm{ms}}}{dx}
=-\tau\frac{d\tau}{dx}\frac{ c}{2{l}}
\end{equation}
for the coupling constants, where
\begin{equation}
\label{derivatives1}
  \begin{array}{cc}
&\frac{d \tau}{d x}= -4kr_{\mathrm{ms}}R_bT_b\sin2kx,  \\
&\frac{d \mu}{d x}= -2kr_{\mathrm{ms}}[2t_{\mathrm{ms}}R_bT_b\cos2kx-r_{\mathrm{ms}}(T_b^2-R_b^2)].  \\
  \end{array}
\end{equation}

We are interested in the values of $g_{\gamma0}$ and $\gamma_{\mathrm{ms}}$ for the position $x$ of the membrane where the dispersive coupling vanishes.
According to (\ref{go1}), this happens when  $\frac{d \mu}{d x}=0$, implying via Eq.~(\ref{derivatives1}) the condition for $x$, which reads
\begin{equation}
\label{cond00}
\cos2kx=r_{\mathrm{ms}}\frac{T_b^2-R_b^2}{2t_{\mathrm{ms}}R_bT_b}.
\end{equation}
Under this condition, according to  Eq.~(\ref{MStau})
\begin{equation}
\label{tau00}
\tau= \frac{T_b^2-R_b^2}{t_{\mathrm{ms}}}.
\end{equation}
For the validity of our calculations, we need $|\tau|\ll 1$, yielding
$$
T_b^2\approx R_b^2\approx\frac{1}{2}
$$
and as a result
\begin{equation}
\label{cond001}
|\cos2kx| =|r_{\mathrm{ms}}\tau|\ll1.
\end{equation}
To be specific, we will work close to the point where $2kx\approx\pi/2$.
Then Eq.~(\ref{derivatives1}) implies
\begin{equation}
\label{tau001}
\frac{\partial \tau}{\partial x}\approx -2kr_{\mathrm{ms}}
\end{equation}
and
\begin{equation}
\label{gg00}
\left| \frac{d\gamma_{\mathrm{ms}}}{dx}\right | =\left|\frac{d\tau^2}{dx}\right |\frac{ c}{2l} \approx 2\frac{ |T_b^2-R_b^2|r_{\mathrm{ms}}}{t_{\mathrm{ms}}}\frac{\omega_c}{l}=2\sqrt{T_{\mathrm{ms}}}r_{\mathrm{ms}}\frac{\omega_c}{l}.
\end{equation}
Equations (\ref{gg1}) and (\ref{gg00}) bring us to Eq.~(\ref{gMSI}) of the main text.

\section{Membrane-at-the-edge system}
\label{MATEapp}
\subsection{Vanishing of the dispersive coupling}
\label{MATEapp2}
For MATE, we are interested in the position of the membrane where the dispersive coupling vanishes.
Solving  the following  well-known resonance equation \cite{jayich2008,dumont2019}
\begin{equation}
\label{fr}
\cos(kl+\varphi_r) = -r_m\cos(2kx-kl),
\end{equation}
we find
\begin{equation}
\label{solution}
2x-l= \frac{1}{k} \left[ \pm \cos^{-1}\left(\frac{\cos(kl+\varphi_r)}{r_m}\right) +2\pi N\right],
\end{equation}
where $N$ is an integer, and calculate $dk/dx$ at the resonance values of $k$, $k=\omega_c/c$:
\begin{equation}
\label{solutionDER}
\left(\frac{dk}{dx}\right)^{-1}=\frac{l}{2k} \left[1-\frac{2x}{l} \pm r_m^{-1}
\sqrt{1+\frac{t_m^2\cos^2(kl-2kx)}{1-\cos^2(kl-2kx)}}
\right].
\end{equation}
Equation (\ref{solutionDER}) implies that the dispersive coupling vanishes, i.e. $d\omega_c/dx=0$, at the resonance wave vector satisfying the following condition
\begin{equation}
\label{vanMIM}
\cos^2(kl-2kx)=1,
\end{equation}
or, alternatively, after some algebra, at
\begin{equation}
\label{vanMIM1}
\cos(2kx+\varphi_r)= -r_m.
\end{equation}
In the case of interest where $r_m$ is close to $1$, Eq.~(\ref{vanMIM1}) implies that $\Phi$ defined by Eq.(\ref{Fi}) is small such that Eq.~(\ref{vanMIM1}) yields
\begin{equation}
\label{Fi1}
-1 +(2\Phi)^2/2 = -1 + t_m^2/2.
\end{equation}
The solution to this equation reads
\begin{equation}
\label{Fi11}
\Phi^2 =  t_m^2/4 =\Phi_0,
\end{equation}
which is the result used in the main text.
\subsection{Condition on $x$ for the enhanced optomechanical performance of  MATE}
\label{MATEapp3}
Let us find the condition on $x$, enabling the enhanced value of dispersive coupling constant of MATE identified in Ref.~\citenum{dumont2019}.
For $\lambda\ll l$ and $x\ll l$, according to Eq.~(\ref{solutionDER}), the maximum modulus of the dispersive coupling constant is reached at $\cos(kl-2kx)=0$ while taking $"-"$ in this formula.
Such a maximum value reads
\begin{equation}
\label{gom}
g_{\omega0}= \frac{d\omega_c}{dx} =\frac{\omega_c}{x+lt_m^2/4}.
\end{equation}
This relation implies that the aformentioned enhanced value of the dispersive coupling constant of MATE, which is equal to $4\omega_c/(lt_m^2)$, corresponds to
\begin{equation}
\label{x1}
x \ll l\frac{t_m^2}{4}
\end{equation}
This brings us to inequality (\ref{x}) of the main text.

\subsection{Dispersive coupling at $\Phi=\Phi_0$}
\label{MATEapp1}
According to Eq.~(\ref{solutionDER}), to evaluate the dispersive coupling constant at $\Phi=\Phi_0$, it suffices to know $\cos^2(kl-2kx)$.
To find it, we note that Eq.~(\ref{fr}) can be rewritten as follows
\begin{equation}
\label{U}
\tan(kl-2kx)=\frac{r_m+\cos(2kx+\varphi_r)}{\sin(2kx+\varphi_r)},
\end{equation}
while, at  $\Phi=\Phi_0$, and $t \ll t_m \ll1$, Eq.~(\ref{van}) implies
\begin{equation}
\label{van11}
\cos(2kx+\varphi_r) = - \frac{2r_m}{1+r_m^2}, \qquad \sin(2kx+\varphi_r) = \pm \frac{1-r_m^2}{1+r_m^2}.
\end{equation}
Combining the above relations we find

\begin{equation}
\label{U1}
\cos(kl-2kx)^2=\frac{1}{1+\tan(kl-2kx)^2}=\frac{1}{1+r_m^2}=1/2,
\end{equation}
leading, for the two modes corresponding to $\pm$ in (\ref{solutionDER}), to the following expressions for the
dispersive coupling constants
\begin{equation}
\label{go+}
g_{\omega0+}= -\frac{d\omega_c}{dx}=-\frac{\omega_c}{l-x+lt_m^2/2}
\end{equation}
and
\begin{equation}
\label{go-}
g_{\omega0-}= \frac{d\omega_c}{dx} =\frac{\omega_c}{x+lt_m^2/2},
\end{equation}
respectively.

The mode exhibiting coupling constant given by Eq.~(\ref{go-}) is relevant to our consideration.
The reason is as follows.
The spectrum of the whole cavity in the $k-x$ plane is, actually, made of the resonance curves of its two parts with small areas of the avoided crossing.
Evidently, the dispersive coupling constant of the resonance curves originating from the resonance curves for the $x$-long part is positive while  the dispersive coupling constant of the resonance curves originating from the resonance curves for the $l-x$-long part is negative.
Addressing $\Phi\ll 1$, we are close to the line given by  equation $\cos(2kx+\varphi_r)=-1$, which is the resonance curve for the $x$-long part.
Thus, we conclude that, for $\Phi\ll 1$, the dispersive coupling constant should be positive as that given by Eq.~(\ref{go-}) is.

Next, in view of condition (\ref{x1}), Eq.~(\ref{go-}) yields
\begin{equation}
\label{go--}
g_{\omega0-}= \frac{d\omega_c}{dx} =\frac{\omega_c}{l}\frac{2}{t_m^2}
\end{equation}
and, finally, combining (\ref{gMO}) and (\ref{go--}) we find
\begin{equation}
\label{gratio}
\left| \frac{g_{\gamma0}}{g_{\omega0-}}\right|= \frac{t^2}{t_m^2} =\frac{T}{2}\ll 1,
\end{equation}
where Eq.(\ref{T1}) is taken into account.
Equation (\ref{gratio}) implies that, at $\Phi=\Phi_0$, MATE is dominated by the dispersive coupling.
\subsection{Applicability of the synthetic mirror approach to MATE}
\label{MATEapp4}
Let us show that under conditions (\ref{x1}) and $\lambda\ll l$, the results for the decay rate and dissipative coupling constant obtained in Sec.\ref{analysis} using the synthetic mirror approach can be  applied to MATE.

According to Ref.~\citenum{dumont2019}, for $t\ll t_m$, the decay rate of MATE reads
\begin{equation}
\label{MIMdec}
\gamma_{\mathrm{mate}} = \frac{ct^2t_m^2/2}{xt_m^2+(l-x)[1+r_m^2+ 2r_m \cos(2kx+\varphi_r)]},
\end{equation}
which can be rewritten as follows
\begin{equation}
\label{MIMdec1}
\gamma_{\mathrm{mate}} = \frac{cT}{2l}\frac{1}{1+A},
\qquad A =\frac{xt_m^2}{1+r_m^2+ 2r_m \cos(2kx+\varphi_r)},
\end{equation}
where $T$ comes from Eq.~(\ref{T}).
In the situation of interest, where $\cos(2kx+\varphi_r)\approx -1$ , in view of (\ref{x1}), $A= 4x/(lt_m^2)\ll 1$ such that the use of (\ref{gamma2}) for the calculation of the  MATE decay rate is justified.

According to Ref.~\citenum{dumont2019}, for $t\ll t_m$, the dissipative coupling constant of MATE reads
\begin{equation}
\label{MIMdecay}
\frac{d\gamma_{\mathrm{mate}}}{dx} =ct^2t_m^2
 \frac{r_m^2+r_m \cos(2kx+\varphi_r) + 2r_mk(L-x)\sin(2kx+\varphi_r) }
 {\{l[1+r_m^2+ 2r_m \cos(2kx+\varphi_r)]-2x[r_m^2+r_m \cos(2kx+\varphi_r)]\}^2 },
\end{equation}
Here, as was shown just above, condition (\ref{x1}) enables dropping of the second term in the denominator while, for $\cos(2kx+\varphi_r)\approx -1$ , the numerator can be rewritten as follows
$$
-t_m^2/2+2kl\sin(2kx+\varphi_r).
$$
In the present text, we discuss MATE for $\Phi\geq\Phi_0$ implying, via (\ref{van11}),  $|\sin(2kx+\varphi_r)|\geq t_m^2/2$ such that the first two  terms in the numerator in (\ref{MIMdecay}) can be dropped if $\lambda\ll l$.
Thus we find
\begin{equation}
\label{MIMdecay1}
\frac{d\gamma_{\mathrm{mate}}}{dx} =\frac{ck}{l}\frac{2t^2t_m^2\sin(2kx+\varphi_r)}{[1+r_m^2+ 2r_m \cos(2kx+\varphi_r)]^2}.
\end{equation}
This relation is consistent with the result given by Eq.~(\ref{der0}) and (\ref{T/x}), which is  obtained using the syntectic mirror approach.
\bibliography{QOwork,NF}
\end{document}